\documentclass[useAMS,usenatbib]{mn2e}

\usepackage{graphicx}
\usepackage{subfig}
\usepackage[T1]{fontenc}
\usepackage{aecompl}

\newcommand{\apleq}{\ensuremath{\stackrel{<}{_\sim}}}
\newcommand{\kap}[1]{Sect.\,\ref{#1}}
\newcommand{\mem}[1]{\ensuremath{\mathrm{ #1}}}
\newcommand{\czw}{\ensuremath{^{12}\mem{C}}}
\newcommand{\nezwa}{\ensuremath{^{20}\mem{Ne}}}
\newcommand{\msun}{\ensuremath{\, M_\odot}}
\newcommand{\apgeq}{\ensuremath{\stackrel{>}{_\sim}}}

\def\aap{\ref@jnl{A\&A}} 
\def\araa{\ref@jnl{ARA\&A}} 
\def\aaps{\ref@jnl{A\&AS}} 
\def\mnras{\ref@jnl{MNRAS}} 
\def\apj{\ref@jnl{ApJ}} 
\def\apjs{\ref@jnl{ApJS}} 
\def\apjl{\ref@jnl{ApJ}} 
\def\aj{\ref@jnl{AJ}} 
\def\memsai{\ref@jnl{Mem.~Soc.~Astron.~Italiana}}
\def\prc{\ref@jnl{Phys.~Rev.~C}} 

\title[Evolution of SN Ia progenitors]{The dependence of the evolution of SN type Ia progenitors on
the C burning rate uncertainty and parameters of convective boundary mixing}
\author[Chen et al.]{Michael C. Chen$^{1}$
, Falk Herwig$^{1,2}$, Pavel A. Denissenkov$^{1,2}$ and Bill Paxton$^3$\\
$^{1}$Department of Physics \& Astronomy, University of Victoria,
       P.O.~Box 1700, STN CSC, Victoria, B.C., V8W~2Y2, Canada\\
$^{2}$The Joint Institute for Nuclear Astrophysics, Notre Dame, IN 46556, USA\\
$^{3}$Kavli Institute for Theoretical Physics and Department of Physics, Kohn Hall, University of California, Santa Barbara, CA 93106, USA}

\begin{document}
\date{Accepted 2013 December 31. Received 2013 December 31; in original form 2013 December 31}
\pagerange{\pageref{firstpage}--\pageref{lastpage}} \pubyear{2013}

\maketitle

\label{firstpage}

%
%

\begin{abstract}
  Evolution of a supernova type Ia progenitor requires formation of a CO white dwarf, which implies a dependence on the C burning rate (CBR). It can also be affected by the recently identified possibility of C flame quenching by convective boundary mixing.  We present first results of our study of the combined effect of these two potential sources of uncertainty on the SN Ia progenitor evolution. We consider the possibility that the CBR is higher than its currently recommended value by as much as a factor of 1000 if unidentified resonances are important, or that it is significantly lower because of the hindrance effect. For stellar models that assume the Schwarzschild boundary for convection, the maximum initial mass for the formation of CO WDs increases from $M_\mathrm{i}\approx 5.5\msun$ for the CBR factor of 1000 to $M_\mathrm{i}\apgeq 7.0\msun$ for the CBR factor of 0.01.  For C-flame quenching models, hybrid C-O-Ne WDs form for a range of initial mass of $\Delta M_\mathrm{i}\approx 1\msun$, which increases a fraction of stars that form WDs capable of igniting C in a thermonuclear runaway.  The most extreme case is found for the CBR factor of $0.1$ that is supported by the hindrance model.  This nuclear physics assumption, combined with C flame quenching, leads to the formation of a hybrid C-O-Ne WD with a mass of $1.3\msun$. Such WDs do not need to accrete much mass to reach the Chandrasekhar limit.

\end{abstract}

\begin{keywords}
methods: numerical --- stars: evolution --- stars: interiors --- stars: AGB and post-AGB --- stars: white dwarfs
\end{keywords}

\section{Introduction}
The expansion rate of the Universe has been determined by using type Ia supernovae (SNe Ia) as standard candles \citep{Perlmutter99}. Pathways for the progenitor evolution of SNe Ia include the single degenerate (SD)
\citep{Whelan73,Nomoto1982a,Iben84,Webbink1984}, double degenerate (DD) \citep{Shen12,Raskin2012}, and double detonation scenarios \citep{Woosley94}. All the three scenarios require the presence of at least one CO white dwarf (WD) in a binary system. The thermonuclear ignition of the CO WD is triggered when its mass reaches the Chandrasekhar limit \citep{Hillebrandt00}. In addition to the double detonation, several sub-Chandrasekhar CO WD scenarios have been proposed in the recent years \citep{Maoz2012}.

In any case, a SN Ia explosion requires first formation of a CO WD. The observed SN Ia rates can be used to constrain the theoretical models of SN Ia progenitor systems \citep{Maoz2012}. Population synthesis models rely on input from the theory of stellar evolution to determine the rates and delay times of SN Ia \citep{Hachisu96,Yungelson00,Han04,Ruiter2009,Mennekens10}. One obviously important input for such models is the initial mass range of CO WDs. The closer the initial mass of the formed CO WD to the Chandrasekhar limit, the sooner it will reach this limit by accreting material either from its non-degenerate companion (in the SD scanario) or from the double-degenerate post-merger disk (in the DD scenario). In standard stellar evolution models, the maximum mass of a CO WD from single star evolution is $\approx 1.05\msun$. In a CO core above this mass limit, carbon is ignited and its burning eventually turns the entire core into an ONe WD that cannot experience a thermonuclear explosion. An alternative possibility has been proposed by \citet{Denissenkov2013} and it will be discussed below. In standard models, however, carbon is ignited in the cores of stars with initial masses between $7.5\msun$ and $9.25\msun$ \citep{Garcia-Berro97,Poelarends08}, and an ONe WD that reaches the Chandrasekhar mass ends up as an electron-capture supernova collapsing into a neutron star \citep{Miyaji80,Jones2013}.

The value of the upper initial mass limit for stars that produce WD progenitors of SNe Ia, $M_\mathrm{up}$, predicted by
stellar evolution simulations depends on several physics assumptions and model parameters, such as stellar rotation and metallicity. For example, M$_\mathrm{up}$ increases with metallicity \citep{Becker79,Meng08}. Another important assumption is 
the carbon (${^{12}\mathrm{C}}+{^{12}\mathrm{C}}$) burning rate (hereafter, CBR). For a larger CBR, C ignites at a lower temperature, hence at a lower core mass corresponding to a lower initial mass. A larger CBR therefore implies a smaller value for $M_\mathrm{up}$ and vice versa. 
Like rates of other important charged particle reactions, the CBR at stellar temperatures is determined by extrapolation of experimental data obtained for a higher energy in the interval between 2 MeV and 6 MeV to the Gamow peak energy of about 1.5 MeV, corresponding to C burning at the temperature of $T = 5\times 10^{8}\mem{K}$ \citep{Barron-Palos06}. Possible resonance \citep{Spillane07} or hindrance \citep{Jiang07} have been speculated to be present near the Gamow energy and, as a result, the uncertainty of the total CBR may be of several orders of magnitude. The impact of different CBRs on the evolution and nucleosynthesis of massive stars \citep{Gasques07, Bennett12, Pignatari13} as well as on the superburst ignition \citep{Cooper2009} have been explored in the literature.

The exact value of $M_\mathrm{up}$ affects population synthesis predictions for SN Ia progenitors in two ways. A larger value of $M_\mathrm{up}$ means that the more massive and shorter lived AGB stars can produce CO WDs, which therefore reduces the delay time of SNe Ia. The delay time is the time taken for SNe Ia to appear after a star formation burst. Besides, a larger value of $M_\mathrm{up}$ increases the total number of progenitors of WDs that can explode as SNe Ia. In the SD scenario, a larger value of $M_\mathrm{up}$ also means that the most massive CO WDs need to accrete less material to reach the Chandrasekhar limit.

In addition to the uncertainy of the CBR, it has recently been shown that the treatment of convective boundaries in CO cores of super-AGB stars may have an important effect on the initial mass range of possible SN Ia progenitors. In the standard model, C burning propagates all the way to the centre after off-centre C ignition and, as a result, the entire CO WD turns into an ONe WD, even when C burning starts far from the centre. \citet{Denissenkov2013} have constructed models of super-AGB stars that ignite C off centre and include a small amount of convective boundary mixing similar in its length scale to those required to explain the observed abundance anomalies in classical novae and in H-deficient post-AGB stars \citep{Werner2006,Denissenkov-nova-2013}. In these models, convective boundary mixing causes the C-flame to stall, which eventually prevents it from reaching the centre. This produces a C-O-Ne hybrid WD at the end of super-AGB star evolution with a significant unburnt CO core surrounded by a thick ONe shell. With a sufficient amount of C still left in the centre some super-AGB stars could be candidates to provide WD progenitors of SNe Ia. 
{\bf Such SNe can probably be distinguished from their more common counterparts resulting from carbon explosion in
CO WDs because the mass-averaged abundance of C in hybrid WDs is significantly smaller and it depends on the initial mass of 
the star. This can potentially explain the observed diversity of SNe Ia in their explosion strength \citep{Waldman07}.}
 If such C-O-Ne hybrid WDs are confirmed as SN Ia progenitors, the value of M$_\mathrm{up}$ may no longer be considered as the largest initial mass ($M_\mathrm{i}$) of a star that can produce a CO WD, but rather as the largest $M_\mathrm{i}$ that can produce a hybrid WD with a sufficiently large CO core to ignite a SN Ia. The addition of the hybrid WDs to the CO WDs will not only increase the number of SN Ia progenitors but also decrease the average delay time of SNe Ia.

The aim of this paper is to investigate the upper initial mass limit for stars that can produce WD progenitors of SNe Ia by taking into account the uncertainties in the CBR in combination with different assumptions about convective boundary mixing, including the possibility of the formation of hybrid WDs. In \kap{sec:method}, we briefly describe our simulation tools before we present the results in \kap{sec:results}. Finally, a discussion is provided in \kap{sec:diss_n_concl}.

\section{Simulation Methods} \label{sec:method}
The stellar evolution simulations performed in this study used the Modules for Experiments in Stellar Evolution (MESA) code version 4442 \citep{Paxton11}. Each stellar evolution simulation begins with a pre-main sequence star model. The mass of the model is first automatically chosen by the code at a value closest to our specified initial mass and then adjusted to the targeted value using the MESA controlled mass gain or loss.  

We have to calculate masses of H- and He-free cores. The boundaries of these cores are defined by the mass coordinates at which the abundances of H and He reach specified low values when being approached from the stellar surface downwards. In this paper, we use the mass fractions of H and He equal to 0.01 and 0.05 for the H- and He-free core boundaries, respectively. These values correspond to the places of the maximum energy production in the H and He burning shells. 

The adopted nuclear network contains reactions of H, He and C burning. Most of the reaction rates used in MESA are taken from Caughlan \& Fowler \citep[CF88,][]{Caughlan88}, and Angulo et al. \citep[NACRE,][]{Angulo99}. We vary the CBR by applying a multiplicative factor (0.01, 0.1, 1, 100 and 1000) to its standard recommended value. The nuclear physics of the \czw+\czw\ rate has been summarized recently by \citet{Pignatari2013} and \citet{Bennett12}, and we refer the reader to their discussions. The possible reduction of the CBR due to the hindrance effect strongly increases when the temperature decreases. The temperature characteristic of C-flame burning is $\log T = 8.9$, for which the hindrance model suggests a reduction of the CBR by a factor of 10 \citep{Jiang07,Gasques07}.  Most carbon burning takes place in super-AGB star models at $T$ between $7.2\times 10^8$ K and $7.6\times 10^8$ K.
At temperatures as low as  $\log T = 8.8$, the hindrance effect may reduce the CBR  by a factor of $\approx 50$. We consider the CBR reduction factors of $0.01$ and $0.1$ to be associated with the hindrance. The smaller factor is probably a bit less than that presently considered as the most reasonable lower limit for the CBR uncertainty range. Hidden resonances may increase the CBR substantially and, following \citet{Pignatari2013}, we consider for them the enhancement factors of $10$, $100$ and $1000$. 

The main goal of our analysis is to estimate the mass limits for the CO, C-O-Ne hybrid and ONe WDs. For each value of the CBR and assumption about convective boundary mixing, a grid of stellar models with initial masses in the range between $5\msun$ and $9\msun$ with a fine mass resolution was computed, so that the transition masses were determined within $0.25\msun$. All the simulation runs have the initial metallicity of $Z=0.01$. The C and O enhanced opacities were enabled. The opacities used by MESA are derived from the Type 1 and 2 OPAL tables \citep{Iglesias93,Iglesias96}, optionally from the OP tables \citep{Seaton05}, and from \citet{Ferguson05}. The initial metallicity of the star is used as the base metallicity for the CO enhanced opacities.

Convective mixing is treated by MESA as a diffusive process with a diffusion coefficient $D_\mathrm{mix} = D_\mathrm{conv}$. The coefficient $D_\mathrm{conv}$ is determined using a mixing length theory module. MESA treats convective overshooting, that we call here convective boundary mixing, as an exponential decay of $D_\mathrm{mix}$ immediately outside a convective zone on a length scale of $f$ pressure scale heights at the Schwarzschild boundary. This should appropriately approximate mixing caused by hydrodynamic instabilities at convective boundaries. For all evolutionary phases preceding C ignition, we use the convective boundary mixing parameter $f = 0.014$. This value has been fitted to the observed terminal age main sequences for a large number of star clusters \citep{Herwig00}. For convective boundaries associated with any nuclear burning other than hydrogen burning, the $f$ parameter is set to be zero to reproduce the standard case that does not produce any C-O-Ne hybrid cores, and $f=0.007$ is adopted for the convective boundary mixing at the bottom of the C-burning convection zone following \citet{Denissenkov2013}.

Special care has been taken to ensure that the C flames are properly resolved. We add mesh resolution constraints to ensure very high spatial resolution with respect to the \czw\ abundance and $T$ profile as described by \citet{Denissenkov2013}, and we reproduce their results. 

Our computations are set to terminate at the first thermal pulse. However, models with convective boundary mixing sometimes become numerically demanding after the end of the second dredge-up in that their core masses do not change much anymore before the first thermal pulse occurs. Given that during this phase the core mass would only change by $\left |\Delta M_{\odot }  \right |\apleq 0.01\msun$, we have decided to skip this phase and report core masses attained by the end of the second dredge-up.

\section{Results}\label{sec:results}
We have performed two simulation sets. The first set was done with no convective boundary mixing during C burning ($f =0$) to reproduce the standard scenario in which the ignited carbon flame never fails in transforming a CO core inside a super-AGB star into an ONe core. The second set used the convective boundary parameter $f = 0.007$, which resulted in hybrid C-O-Ne WDs.
 All the other parameters in each set, except the initial mass and the CBR factor, had the same values.

\subsection{The evolution of stars without convective boundary mixing --- the standard case} \label{subSec:noOVS}

\begin{figure*}
	\centering
	\subfloat[]{\includegraphics[width=0.48\textwidth]{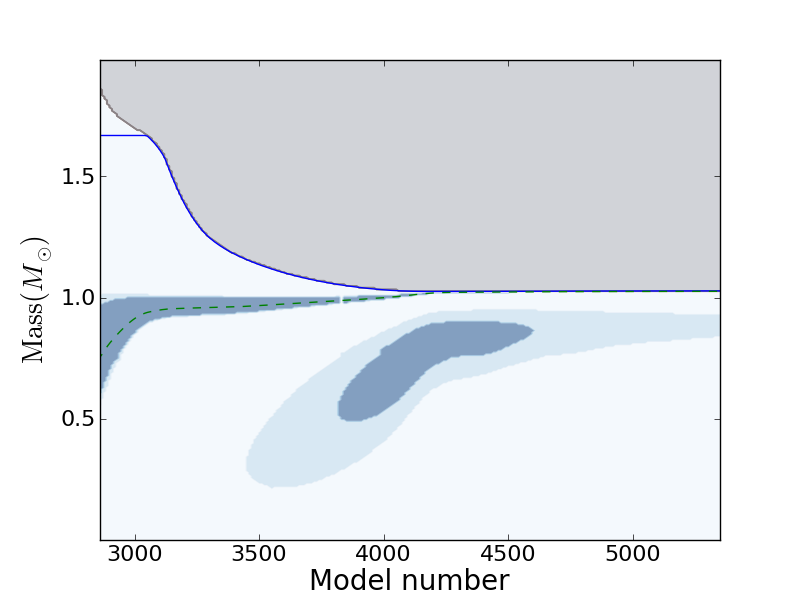}
		\label{fig:KipCO_noOVS}}
	\subfloat[]{\includegraphics[width=0.48\textwidth]{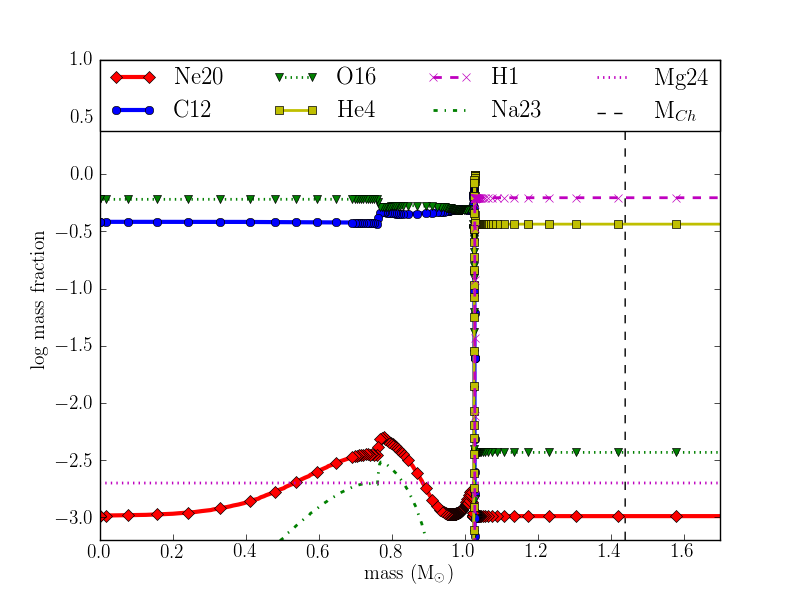}
		\label{fig:abndCO_noOVS}}\\

	\subfloat[]{\includegraphics[width=0.48\textwidth]{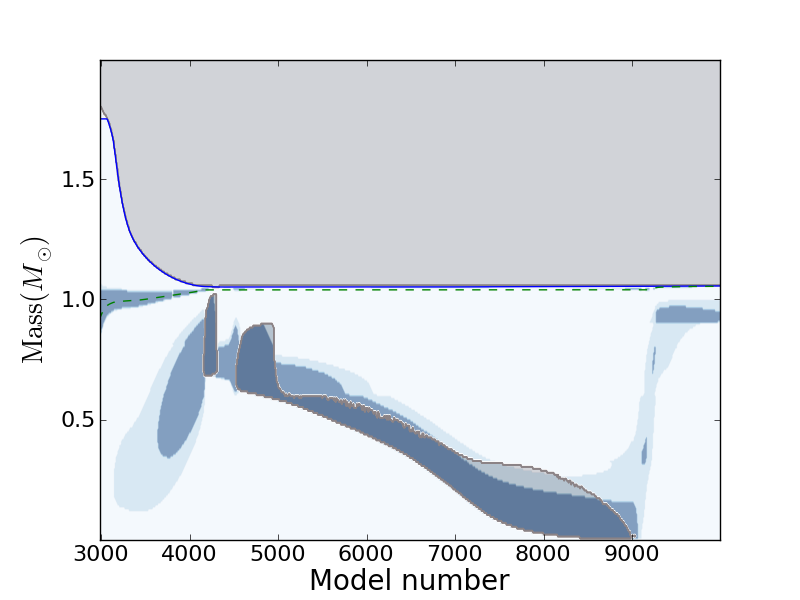}
		\label{fig:kipONe_noOVS}}
	\subfloat[]{\includegraphics[width=0.48\textwidth]{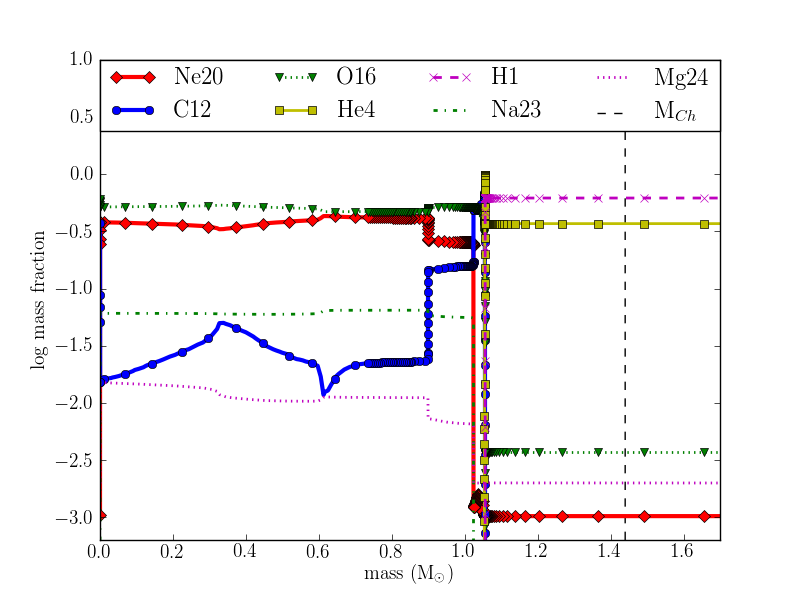}
		\label{fig:abndONe_noOVS}}\\
	\caption{Left: Kippenhahn diagrams for C burning. Right: abundance profiles at the end of simulations. Top row: $6.5\msun$, and bottom: $6.75\msun$. The standard CBR and no convective boundary mixing in the helium-free core. The grey areas in the Kippenhahn diagrams are convective zones, while the different shades of blue mark the levels and locations of energy production.
{\bf The lighter and darker shades of blue respectively mark the regions with energy production rates between 50 and 1000 
erg\,g$^{-1}$\,s$^{-1}$ and above 1000 erg\,g$^{-1}$\,s$^{-1}$.}}
\label{fig:kipabuf0}
\end{figure*}

Like in the second set, convective boundary mixing was modeled with $f =0.007$ at the interface between the helium core/shell and hydrogen envelope, while everywhere else we used $f = 0.014$ to be in accord with an observationally constrained stellar evolutionary model prior to the start of carbon burning \citep{Herwig00, Werner2006}. Our results are here briefly compared with the investigation of the C flame propagation in super-AGB stars by \citet{Siess06}. He computed models that did not take into account any mixing mechanisms beyond the classical convection defined by the Schwarzschild criterion. The Kippenhahn diagrams for two of our stellar evolution runs that are representative for the cases with similar initial mass that produced CO cores and ONe cores are shown in Fig.~\ref{fig:kipabuf0} with their respective chemical abundance profiles taken at the end of simulation. 

Like in Siess' study, whenever carbon is ignited in the core of a super-AGB star, its burning is either too week to propagate any distance and therefore it is quickly quenched, or it is strong enough to cause several episodes of convection and to propagate all the way to the centre, thus burning most of the carbon into neon. There is not a single case in which the CO core is left partially burnt. How the C flame usually propagates in the latter case is seen in Fig.~\ref{fig:kipabuf0}c. The deflagration wave of the C flame usually has an intermittent character. The consecutive carbon burning episodes usually start with a carbon flash followed by one or more deflagrations ignited below the first flash. These series of deflagrations eventually lead to the propagation of the carbon flame to the centre of the core.

Since Siess did not include any main-sequence convective core overshooting in his simulations, the least massive star that was able to ignite carbon in his simulations was more massive than its counterpart in our investigation. The difference between their initial masses is $\sim 2\msun$ which is in good agreement with what can be found in the literature \citep{Bertelli1985, Bressan1993, Eldridge2004b, Poelarends08}. Siess also used a metallicity slightly different from that used in our models.

\begin{figure}
	\includegraphics[width=0.5\textwidth]{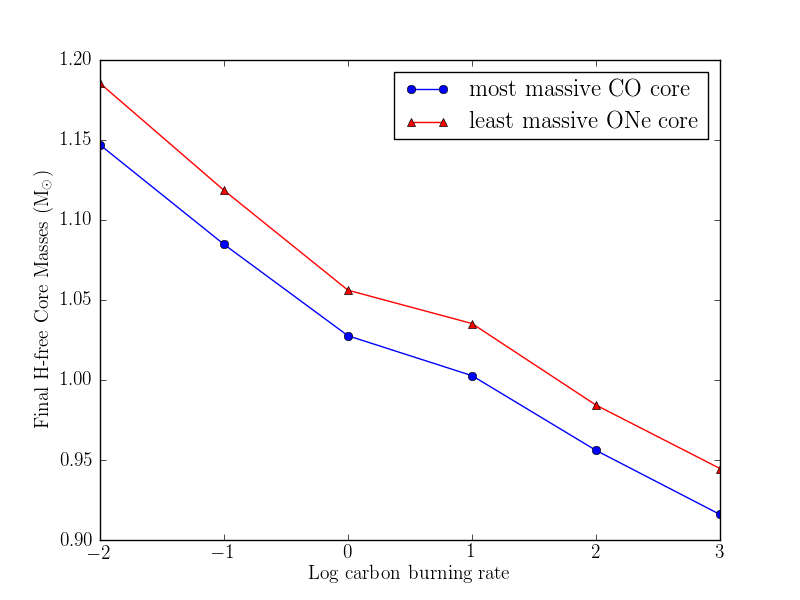}
	\includegraphics[width=0.5\textwidth]{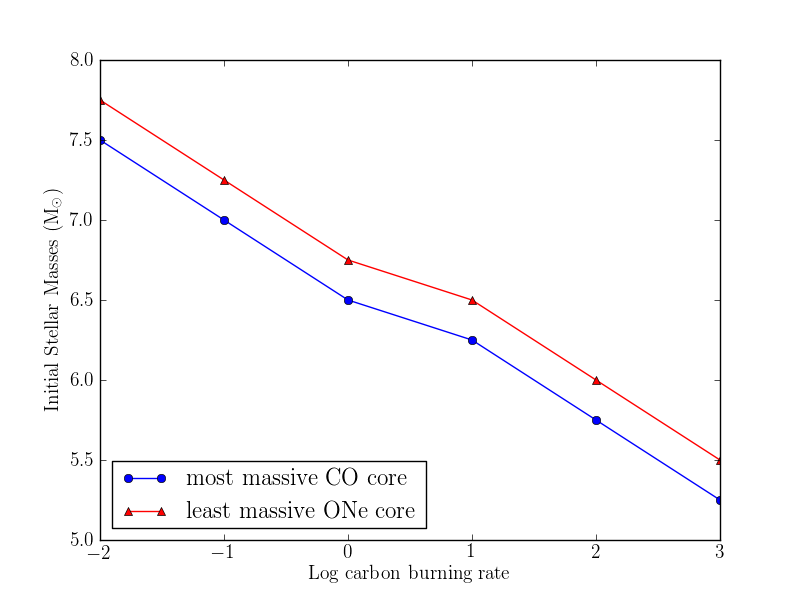}
    \caption{The highest masses of the CO cores and the lowest masses of the ONe cores (top), and the initial masses of the stars that produce the highest CO core masses and the lowest ONe core masses (bottom), all plotted against the carbon burning rate (CBR) factor.
\label{fig:ccM}}
\end{figure}
To illustrate the effect of carbon burning rate (CBR) on the most massive AGB core that can become a SN Ia progenitor, the initial masses and the core masses indicating the transition between the formation of CO and ONe cores are shown in Fig.~\ref{fig:ccM}. The upper limit of the CO core increases as the CBR decreases. Interestingly, the core mass boundary that marks the CO to ONe transition appears nearly linear with respect to the logarithm of the CBR. The actual CO to ONe transition would lie between the two mass limits seen in the figure, making the mass limits the upper and lower bounds of the uncertainty. 
\begin{figure}
	\includegraphics[width=0.5\textwidth]{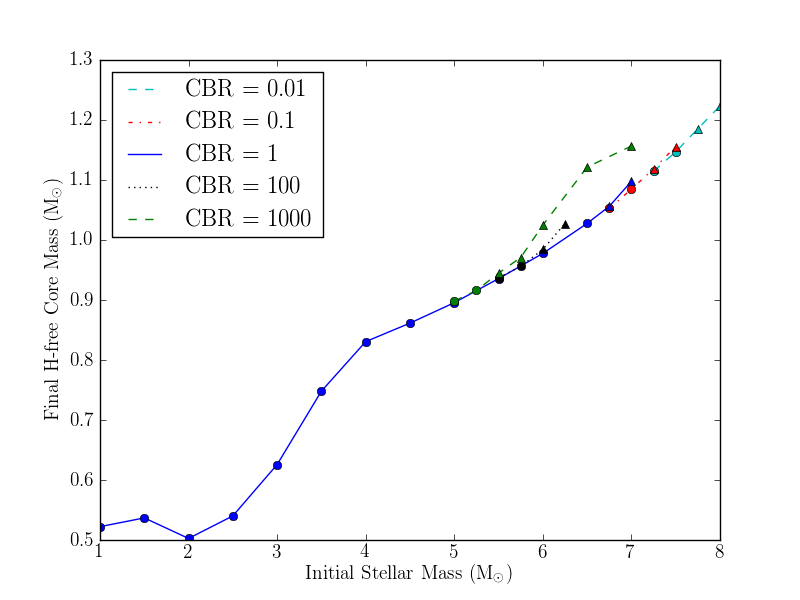}
    \caption{The initial to H-free core mass relation for different CBRs. Circles, squares, and triangles represent the CO, hybrid, and ONe cores, respectively.
\label{fig:MiMf_noOVS}}
\end{figure}

To study the initial final mass relations, (IFMRs), the initial masses from these runs are plotted against the final masses of the hydrogen free cores in Fig.\, \ref{fig:MiMf_noOVS}. {\bf The H-free core of an AGB star consists of a He-free core surrounded by
a He shell whose mass is negligible. Therefore, following \cite{Becker79}, we consider the mass of the H-free core as a main result
of our calculations}. In general, the final core mass $M_\mathrm{f}$ increases with the initial stellar mass $M_\mathrm{i}$. Stars with $M_\mathrm{i}$ less than $2.5\msun$, however, have a roughly constant $M_\mathrm{f}$. This relation is valid down to the lowest $M_\mathrm{i}$ that was modelled, i.e. $1.0\msun$. As expected, the IFMR only has a dependency on the CBR in the mass regime where ONe cores are produced as the result of carbon burning. The $M_\mathrm{f}$ of any given ONe core has always been found to be higher than their counterparts produced from the same $M_\mathrm{i}$ but with a lower CBR.

\subsection{C-shell burning with convective boundary mixing --- formation of the hybrid C-O-Ne WDs} \label{subSec: overshooting}

\begin{figure*}
\includegraphics[width=0.48\textwidth]{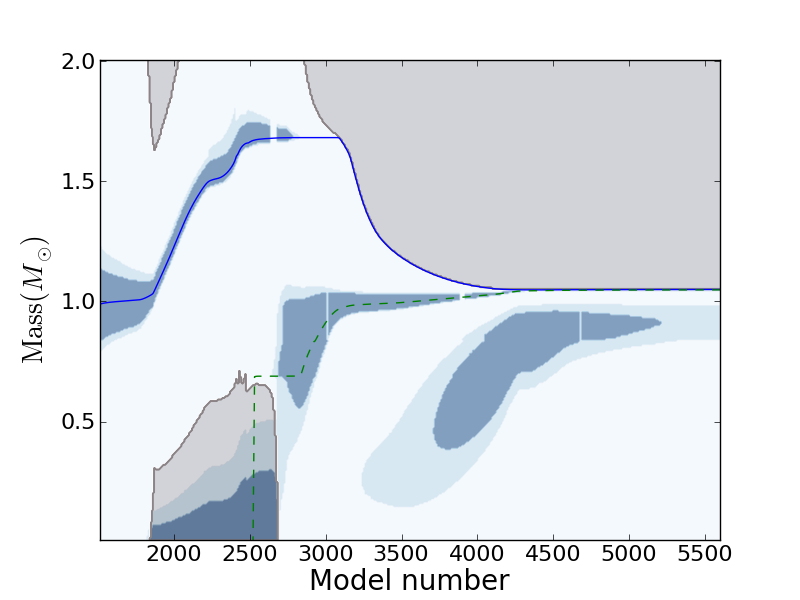}
\includegraphics[width=0.48\textwidth]{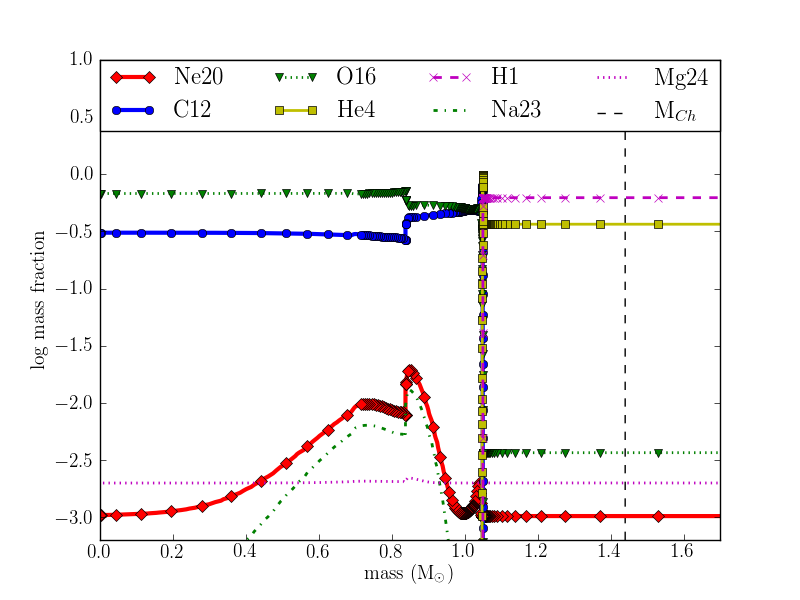}
\includegraphics[width=0.48\textwidth]{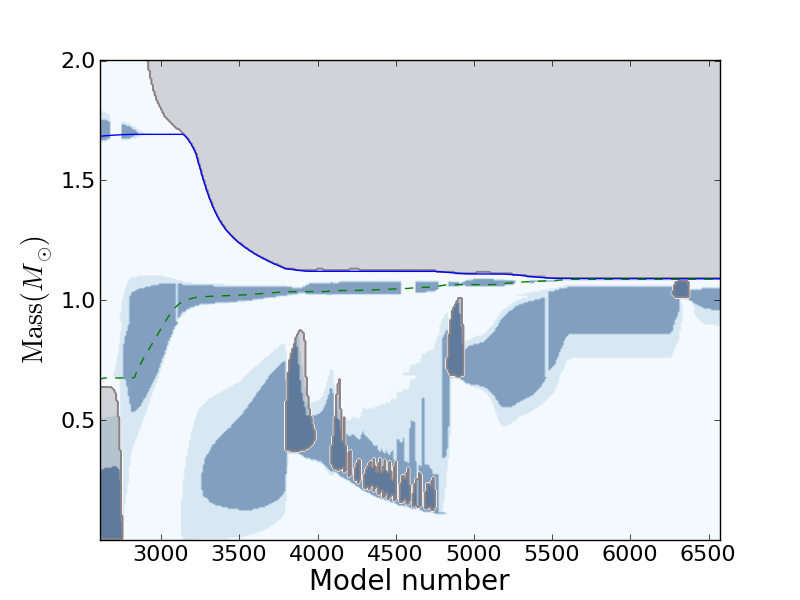}
\includegraphics[width=0.48\textwidth]{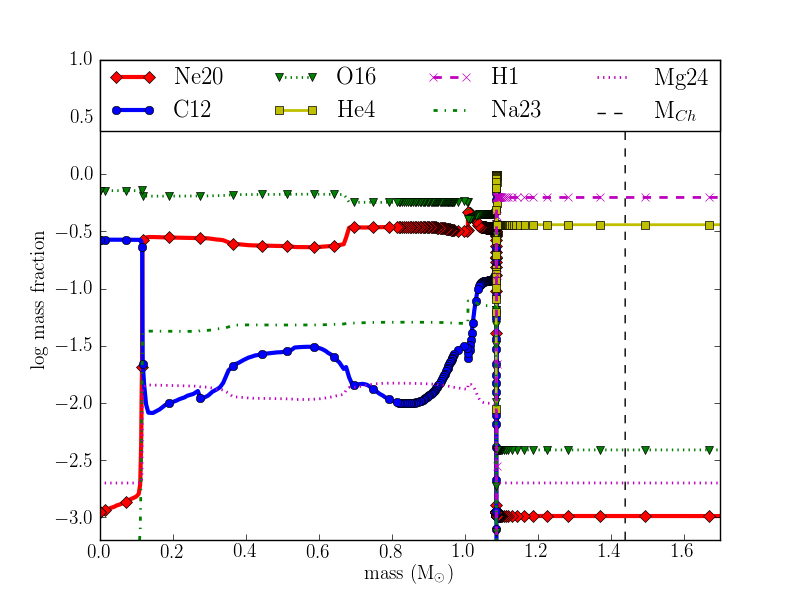}
\includegraphics[width=0.48\textwidth]{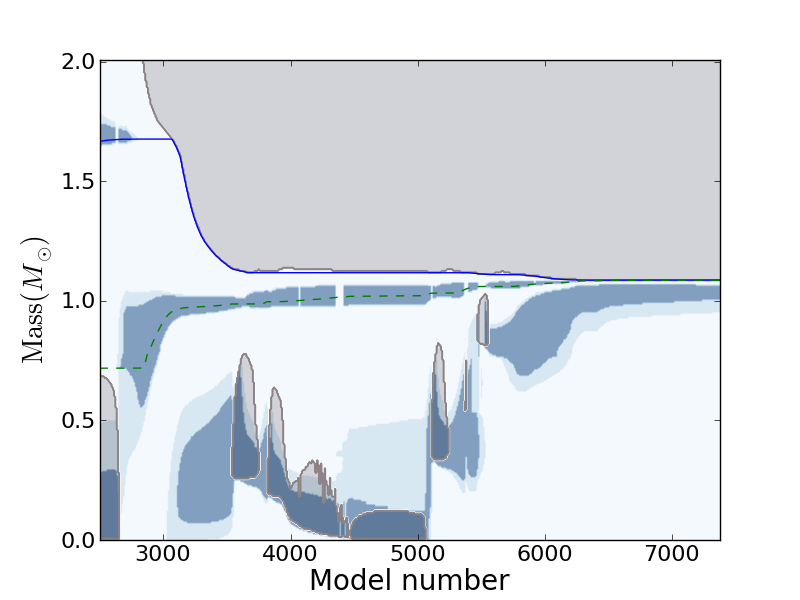}
\includegraphics[width=0.48\textwidth]{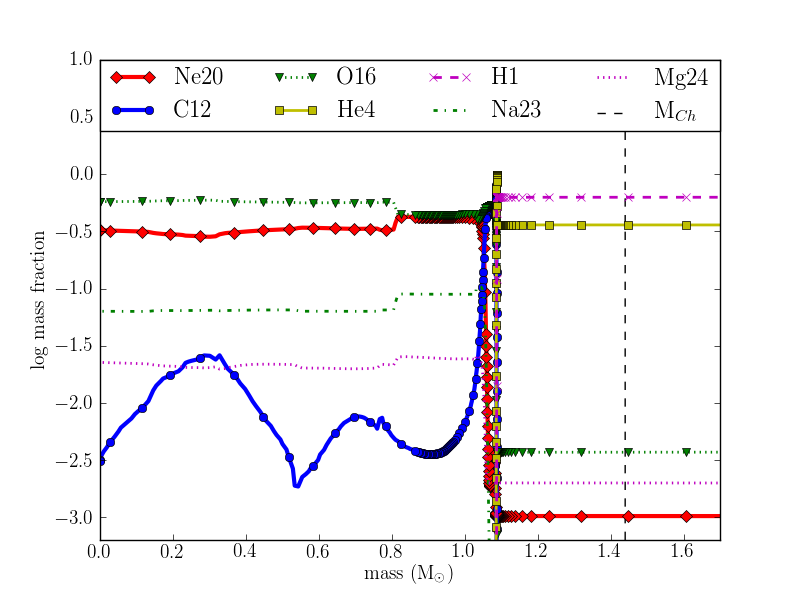}
\caption{\label{fig:withOV} 
Similar to Fig.\,\ref{fig:kipabuf0}; all three cases have an initial mass of $6.5\msun$ and convective boundary mixing for C shell burning. Shown are three runs with CBR factors of $1$, $10$, and $100$ from top to bottom. These stellar models with the same initial mass illustrate the 3 different types of cores that can be produced depending on the CBR. The shown models lie on the black dashed line seen in Fig.\,\ref{fig:ccMconv}. }
\end{figure*}

The second set of our simulations models convective boundary mixing with $f=0.007$ at the bottom of convective C-burning, but otherwise it is the same as the standard case presented above. Our models cover the complete initial mass range for hybrid C-O-Ne WDs. The transitions between the CO and ONe cores as a function of increasing $M_\mathrm{i}$ are occupied by hybrid C-O-Ne cores in this case. The Kippenhahn diagrams of the stellar evolution that led to the creation of CO, hybrid, and ONe cores are shown in Fig.\,\ref{fig:withOV}.

The case with a CBR factor of $1$ shows a weak ignition of C burning that however does not develop into a flame. C burning stops before any substantial amount of \czw\ has been burnt. A small pocket of \nezwa\ remains in the core. However, this case would result in a normal CO WD. Both its Kippenhahn diagram and its abundance profile resemble those of its counterpart with no convective boundary mixing (see Figs.~\ref{fig:KipCO_noOVS} and \ref{fig:abndCO_noOVS}). 

The middle panel shows the case with a CBR factor of 10. This case results in
a hybrid C-O-Ne core as described by \citet{Denissenkov2013}. There appears to exist a continuum of CO core masses that remain unburnt in the centre of hybrid cores. We have defined a hybrid core preliminarily as a core that has a $^{12}$C rich centre which occupies more than 1$\%$ of the total core mass and is surrounded by a $^{20}$Ne rich shell. The central $^{12}$C abundance needs to be at least a hundred times larger than $^{20}$Ne by mass fraction in order to be considered ``rich''. These hybrid cores often have a small unburnt carbon shell lying on top of the neon shell as well.

The hybrid core (middle panel in Fig.\,\ref{fig:withOV}) shows that a mild C burning is first ignited near the mass coordinate of $0.3\msun$ and then propagates upwards until the first convective burning is initiated. The convective zone then disappears without any significant migration followed by a continuous C flame that propagates toward the centre. Short bursts of convective burnings turn on and off as the conductive flame moves towards the centre. The flame dies out before reaching the centre. Very mild burning then migrates upwards to the helium shell and is stopped short of reaching it. Judging from the step-like increase in the $^{20}$Ne abundance as it moves towards the larger mass coordinates, the short episodes of convective burning appear to be more efficient further away from the centre than when they are closer to it. This can be explained by the increase in neutrino cooling as one moves towards the centre, making carbon burning less sustainable near the centre.

Finally, the bottom panel shows the case of a CBR factor of $100$. The abundance profiles are representative of the ONe cores found in the run with a high abundance of $^{20}$Ne and a low abundance of $^{12}$C throughout the helium free core. The evolution shows that convective C burning started off but fairly close to the centre and it was able to propagate all the way to the centre with minor interruptions. The convective burning zone then moved upwards, creating three subsequent episodes of convective burning before it approached the helium shell.

\begin{figure}
	\includegraphics[width=0.5\textwidth]{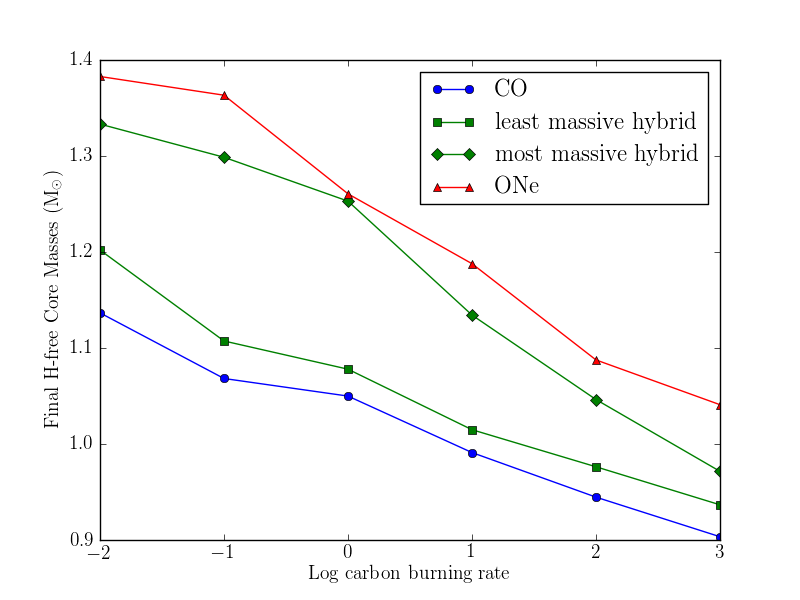}
	\includegraphics[width=0.5\textwidth]{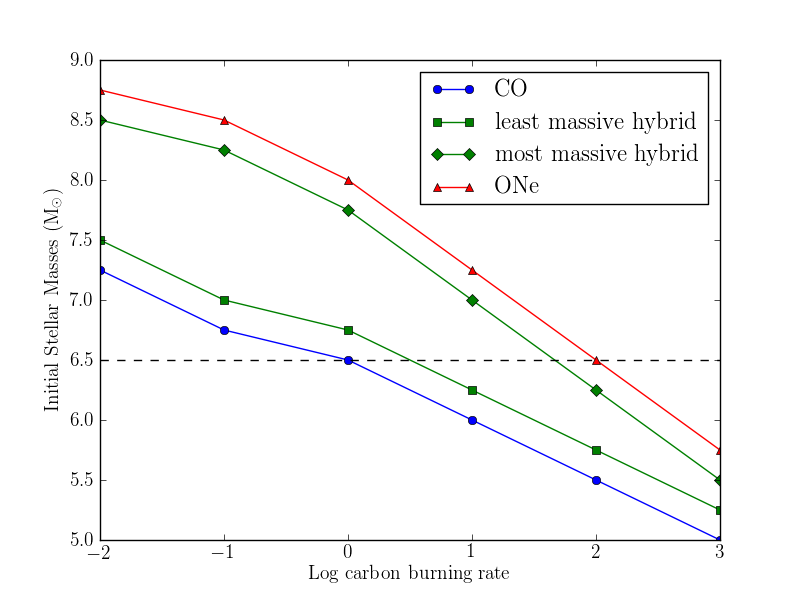}
    \caption{Same as in Fig.\,\ref{fig:ccM}, but for models with convective boundary mixing for C burning. The mass range between CO and ONe WDs is occupied by C-O-Ne hybrid core models.
\label{fig:ccMconv}}
\end{figure}
The highest masses of the CO and hybrid cores, as well as the lowest masses of the hybrid and ONe cores are plotted against the logarithm of the CBR factor in Fig.\,\ref{fig:ccMconv}. As a general trend, the final core mass limits decrease with the increase of the CBR. The range of mass in which hybrid cores can exist decreases as the CBR increases. The final core mass of the most massive hybrid core and the least massive ONe core evolved from the standard CBR usually appear to be close to each other. No anomalies, however, were found in the runs that produced those cores. The initial masses $M_\mathrm{i}$ responsible for the mass limits seen in Fig.~\ref{fig:ccMconv} are plotted against the CBR factor in Fig.\,\ref{fig:ccMconv}. As mentioned earlier in \kap{sec:method}, the transitional initial mass was resolved down to $0.25\msun$ in this study. Much like their corresponding final masses $M_\mathrm{f}$, these initial stellar masses decrease with an increase in the CBR. The range of initial masses that produce the hybrids also decrease as the CBR increases.
\begin{figure}
	\includegraphics[width=0.5\textwidth]{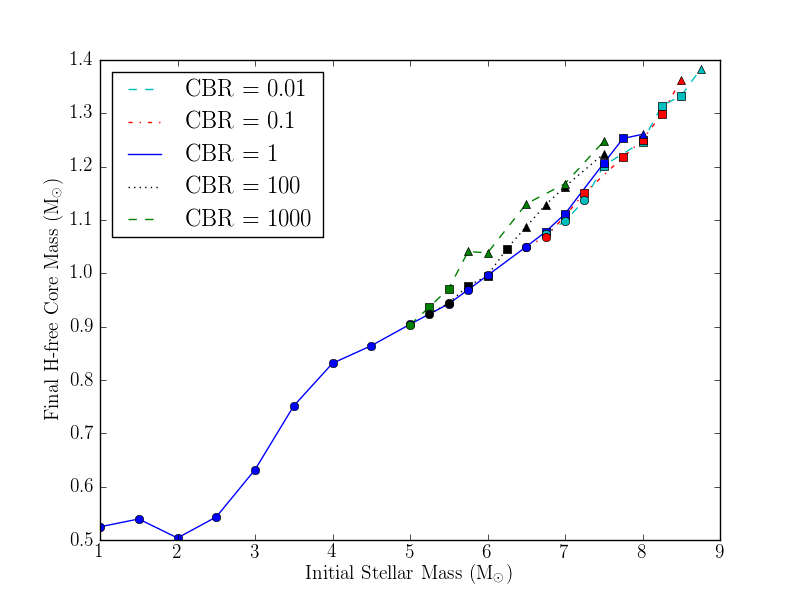}
    \caption{Same as in Fig.\,\ref{fig:MiMf_noOVS}, but for models with convective boundary mixing for C burning.
\label{fig:MiMf}}
\end{figure}

The IFMR is shown in Fig.\,\ref{fig:MiMf}. Aside from the IFMR that branched up due to the earlier offset of carbon burning caused by the higher CBR, the behaviour of the IFMR here is nearly identical to that of its counterpart without convective boundary mixing in its helium-free core. The behaviour of the IFMR that branched off due to carbon burning is also very similar to that of its counterpart without convective boundary mixing. They do, however, branch off a little earlier, most noticeable in the runs with the CBR being 1000 times that of the standard rate.

\section{Discussion and Conclusion}\label{sec:diss_n_concl}
The possibility of the carbon flame being inhibited from propagating to the centre of a super-AGB star was first suggested by Siess in his study of the effects of thermohaline mixing on C-flames \citep{Siess09}, however, without considering the  implications on the WD composition. \citet{Denissenkov2013} demonstrated that thermohaline diffusion coefficients based on hydrodynamic simulations imply mixing characteristics that do not quench the C flame. They also showed that convective boundary mixing should instead quench the C flame and lead to the formation of C-O-Ne hybrid WDs. In light of this new development, the simulations performed in this study were divided into two groups. One group without convective boundary mixing explores the effect of C burning rate (CBR) uncertainties in the more familiar C flame regime of evolution for super-AGB star cores. In the other group, we adopted convective boundary mixing for the C shell convection boundaries. Boundary mixing outside of the helium-free core, however, was left in place to have identical evolution prior to possible C ignition.

Consistent with the earlier studies of C burning that did not account for convective boundary mixing or thermohaline mixing \citep{Garcia-Berro97, Siess06}, all C burning convection zones in our simulations develop flames that propagate to the centre of the core in the absence of convective boundary mixing. Mild C burning that does not initiate convection can neither propagate to a different location nor consume very much carbon. The composition of a CO core that experiences this type of burning remains essentially unchanged. No hybrid core was produced in the runs without convective boundary mixing. Thus, there exists a sharp boundary between the most massive CO core and the least massive ONe core for a given CBR (Fig.\,\ref{fig:ccM}). The initial mass of the star that hosts these cores is anti-correlated with the CBR, as expected.

In the runs with convective boundary mixing the carbon flame quenching was found to be present in all the cores that were just massive enough to ignite convective carbon burning, regardless of their CBR. The flame quenching, however, can be overcome for higher initial and core masses of super-AGB stars. Incomplete burning resulting from the flame quenching has thus created a class of hybrid cores that has an unburnt CO-rich core, surrounded by an ONe shell. The core masses of these hybrids are those of the lower mass ONe cores seen in the no convective boundary mixing case.

The most important conclusion of our investigation with regard to the question of SN Ia progenitors is twofold, depending on what level of convective boundary mixing is appropriate for C burning convection. If convective boundary mixing is zero, the hindrance model would imply that the initial mass for the formation of CO WDs is slightly increased and CO core masses could be up to $1.1\msun$. The more interesting implication however is the possibility of unknown resonances that would increase the C burning rate and may lead to a decrease of the maximum initial mass to form CO WDs and limit the maximum CO WD core mass to $\approx 0.93\msun$.

For the case in which a small amount of convective boundary mixing leads to the formation of hybrid C-O-Ne WDs the implications of the CBR uncertainties are more significant. If C-O-Ne WDs can ignite a thermonuclear runaway, the mass range of SN Ia progenitor WDs could increase significantly. Hybrid WDs that could form if the reduced hindrance CBR is appropriate could be as large as $\approx 1.3\msun$. Such a large WD mass would make reaching the Chandrasekhar limit much easier, since only a small amount of mass has to be accreted. In addition, the maximum initial mass in this case could be just in excess of $8\msun$. This would imply a significantly shorter SN Ia delay time compared to the standard case.

\section*{Acknowledgments}
This research has been supported by the National Science
Foundation under grants PHY 11-25915 and AST 11-09174. This project was
also supported by JINA (NSF grant PHY 08-22648). FH acknowledges funding
from NSERC through a Discovery Grant.


\label{lastpage}

\end{document}